\documentstyle[preprint,aps,tighten]{revtex}
\begin{document}
\draft

\title{Geocentrism re-examined}

\author{Jeremy Goodman\cite{byline}}
\address{Princeton University Observatory, Peyton Hall,
Princeton, NJ 08544}
\date{15 March 1995}
\maketitle

\begin{abstract}
Observations show that the universe is nearly isotropic on very
large scales.
It is much more difficult to show that the universe is radially
homogeneous---that is, independent of distance from us---or
equivalently, that the universe is isotropic about distant points.
This is usually taken as an axiom, since otherwise we would occupy
a special position.
Here we consider several empirical arguments for radial
homogeneity, all of them based on the
cosmic microwave background (CMB).
We assume that physical laws are uniform, but we suppose
that structure on very large scales may not be.
The tightest limits for inhomogeneity on the scale of the horizon
appear to be of order ten percent.
These involve observations of
the Sunyaev-Zel'dovich effect in clusters of galaxies,
excitation of low-energy atomic transitions, and the accurately
thermal spectrum of the CMB.
Weaker limits from primordial nucleosynthesis are discussed
briefly.
\end{abstract}


\narrowtext

\section{Introduction}
\label{sec:intro}

Homogeneity and isotropy are independent cosmological assumptions.
General relativity allows homogeneous but anisotropic universes
({\it e.g.} \cite{HE}), and also
spherically symmetric but inhomogeneous ones
({\it e.g.} \cite{Tolman}).
If the universe is isotropic around two or more distinct
points, however, then it must be homogeneous.

Homogeneity is more fundamental and powerful than isotropy but
also more difficult to verify.
Homogeneity allows local measurements to be applied to the whole
universe; and conversely it allows observations of high-redshift
regions to constrain the history of the local volume.
Without homogeneity, modern cosmology would be very difficult.
Nevertheless, inhomogeneous models are occasionally proposed.
For example,  it has been suggested recently that
spherically symmetric, inhomogenous universes are a natural
consequence of inflation \cite{Linde:94}.
Such speculations, though unorthodox, demonstrate that
homogeneity is not yet fully established.

By contrast, the {\sl isotropy} of the universe on large scales
{\it is} well established.
It is supported by deep, wide-angle surveys of radio
\cite{BWE} and infrared \cite{Setal:92a} sources.
Results from the Cosmic Background Explorer satellite (COBE)
show that the temperature of the microwave background (CMB)
deviates slightly from isotropy, but only
at the level $(\Delta T/T)_{\rm rms} \approx 1.1\times 10^{-5}$
on angular scales $\ge 10^{\circ}$,
apart from a dipole pattern that is conventionally attributed to
the peculiar velocity of the Sun and the Galaxy \cite{Be:94}.

To the extent that the universe is isotropic, it can be
inhomogeneous only if it is symmetric around ourselves.
We therefore ask
whether present observations permit large-scale {\it radial}
inhomogeneity, and if so,  what future measurements might detect
or exclude it.

Galaxy counts against redshift or magnitude
are consistent with a homogeneous, uniformly-populated universe
in the redshift range $0.03\le z\le 0.3$, although statistical
fluctuations associated with structure on scales $\le 30 h^{-1}
\text{Mpc}$, due perhaps to the relatively narrow fields
surveyed, make strong
limits dificult to obtain \cite{KK:92}.
By $z\approx 0.4$, counts in the blue are already
discrepant, perhaps because of
rapid evolution among fainter galaxies \cite{LCG:91}.
Even conservative models predict significant
evolutionary effects on near-infrared counts by $z\sim 1$,
and the model
parameters, though easily adjusted to fit the data,
are poorly constrained {\it a priori} \cite{CLG:90}.
This is typical of the
evolutionary uncertainties that for decades have prevented the use
of galaxies and other beacons to determine cosmic geometry, even
when homogeneity is assumed, because noneuclidean effects are
strong
only for objects so distant as to be seen when the universe was
much younger than it is now.

There may exist ``standard candles'' at $z\gtrsim 1$, such as
Type I supernovae \cite{BT:92}.
Among homogeneous Friedmann models, unfortunately, the shape of the
magnitude-redshift relation for standard candles already depends on two
parameters: the density parameter, $\Omega$, and the cosmological
constant, $\Lambda$.
Only superb data will permit one to fit for a
third parameter and thereby constrain the homogeneity of the
universe on the scale of the present horizon.
Similar remarks apply to more recently-proposed cosmological tests,
such as the use of gravitational lenses to determine the dependence
of angular-diameter distance on redshift \cite{N:91}.

The prospects for constraining the homogeneity
of the CMB are better.
In Sec.\ \ref{sec:therm} we discuss two observational tests that
are sensitive to radial inhomogeneity of the CMB in first order.
Both of these involve measurement of the angle-averaged temperature
of the CMB seen by a distant object, either through scattering or
molecular absorption.
Using recent measurements, we can limit generic
radial inhomogeneities to $\lesssim 10\%$.
We then show in Sec.\ \ref{sec:scatt} that if one assumes
a substantial fraction of all baryons to reside in an ionized
intergalactic medium, then
the accurately thermal nature of the CMB spectrum
provides another $\sim 10\%$ limit, due to second-order
effects of scattering.
In Sec.\ \ref{sec:scatt} we compare these limits with an
argument for homogeneity based on light-element nucleosynthesis.

\section{Thermometers at moderate redshift}
\label{sec:therm}

The CMB is homogeneous if it is isotropic around distinct points.
Imagine therefore that one is provided with a
mirror at a cosmological distance, and that the mirror is
tilted at some angle to the line of sight.
If the universe is isotropic around the distant mirror and the
mirror has negligible peculiar velocity,
then the CMB spectrum seen in the mirror is the same as that seen
directly along unobstructed lines of sight.
If the universe is not homogeneous, then it cannot be isotropic
around both us and the mirror, so the mirror spectrum will
generally differ from the direct spectrum.

To elaborate this idea, imagine that the mirror is half silvered,
so that it reflects a fraction $f$ of the radiation and transmits
the rest (this ideal mirror does not absorb).
Then the spectrum seen in the mirror is
\begin{eqnarray}
\label{eq:mirror}
I_\nu &=& (1-f)B_\nu(T_0) + f B_\nu(T_{\text{r}}) \nonumber \\
&\approx &
B_\nu[(1-f)T_0+ fT_{\text{r}}] ~+~O(\Delta T^2),
\end{eqnarray}
where $B_\nu(T_0)$ and $B_\nu(T_{\text{r}})$ are the direct
and reflected spectra.
We have assumed that the spectrum in any single direction is
thermal.
The combined spectrum is not, unless
$T_{\text{r}}=T_0$, but it
can be approximated by a thermal spectrum to first order in
$T_{\text{r}}-T_0$.

Electron scattering serves as such a mirror.
One requires a cluster of galaxies at redshift
$z_{\text{cl}}\sim 1$ with a nonnegligible
electron-scattering optical depth, $\tau$.
If the cluster fills the telescope beam, the observed spectum
summed over polarizations is, for $\tau\ll 1$,
\begin{eqnarray}
\label{eq:radtrans}
I_\nu^{\rm obs} = (1-\tau)B_\nu(T_0) &+&\tau\int
\case{3}{4}(1+\cos^2\psi) I'_\nu (\Omega)\frac{d^2\Omega}{4\pi}
\nonumber \\
&+& y\,\nu^4\frac{\partial}{\partial\nu}\frac{1}{\nu^2}
\frac{\partial}{\partial\nu} B_\nu(T_0),
\end{eqnarray}
where $B_\nu(T_0)$ is the unscattered thermal spectrum, obtained
from other lines of sight;
$(1+z_{\rm cl})^3 I'_\nu(\Omega)$ is the
specific intensity at the cluster in the direction $\Omega$;
and $\psi$ is the scattering angle between
this direction and the line of sight.
The factor of $(1+\cos^2\psi)$ expresses the angular dependence of
electron scattering, summed over polarizations.
Similar formulae hold for the individual polarizations,
with integrands  depending differently on the scattering angles.
The third term on the right of Eq.\ (\ref{eq:radtrans})
is the Sunyaev-Zel'dovich distortion due to the finite
temperature of the electrons ($T_e\gg T_0$):
$y\equiv \tau k_B T_e/m_e c^2$ \cite{SZ:72}.
Since the first and last term have a known dependence on frequency,
multifrequency observations can be used to constrain the middle
term.

Even in a homogeneous universe, a radial peculiar velocity $v_r$
produces a distortion equivalent to the middle term in
Eq.\ (\ref{eq:radtrans}).
To first order, the anistropic part of $I'_\nu(\Omega)$ is
then a dipole pattern of amplitude $v_r/c$ relative to the monopole
\cite{SZ:80}.
The angular average of this dipole is zero in the cluster rest
frame, so that the cluster sees the same average temperature
as it would if $v_r=0$.
However, an observer at rest with respect to the CMB who views the
cluster along direction $\hat n$ sees the scattered photons to have
suffered a decrease in their energy by a factor of
$(1 -\vec v\cdot\hat n/c)$ on average.
Thus, the conventional interpretation of any non-$y$-type
distortion  would be that the cluster has a peculiar velocity.
In a homogeneous universe, the sign of $v_r$ should be random on
very large scales.
In a radially inhomogeneous universe, $v_r$ will have a
trend with systematically with redshift, and
$\langle v_r\rangle\to 0$ as $z\to 0$ because of spherical
symmetry.
Sec.\ref{sec:scatt} calculates these effects explicitly in
linear theory for a universe close to an Einstein-de Sitter model.

The Sunyaev-Zel'dovich effect has now been measured in several
clusters out to $z\sim 0.2$
in the Rayleigh-Jeans part of the spectrum, where positive
peculiar velocities can not be distinguished from the $y$
distortion \cite{SZRJ}.
Successful results near the Wien peak been reported recently
\cite{SZW}.
The temperature decrements are 1-2~mK with one-sigma errors
$\sim$~10-30\%.
No {\it increments} have been reported, such as might be
produced by a large negative $v_r$.

Lacking multifrequency data for individual clusters, we may
derive a limit on $v_r$ and hence on radial inhomogeneity by
the following argument.
Assuming homogeneity and neglecting $v_r$,
several groups have combined SZ measurements
with X-ray data to derive Hubble's constant, $H_0$
\cite{H0meas}.
The result scales as $H_0\propto \Sigma_X T_e^{5/2}/y^2$,
where $\Sigma_X$ is the Xray surface
brightness at energies $<k_B T_e$.
The electron temperature, $T_e$, can be estimated
directly from the Xray spectrum.
Although the results are smaller than some local
estimates of $H_0$ \cite{Freedman}, they fall within a factor
$\sim 2$, which indicates that peculiar velocities have not altered the
estimates of $y$ by more than $\sim\sqrt{2}$.
Hence
\begin{equation}
\frac{y_{\text{true}}}{\sqrt{2}}<y_{\text{true}}
+\frac{v_r\tau}{2c}<\sqrt{2}y_{\text{true}},
\end{equation}
\begin{equation}
-0.012<\frac{v_r}{c}< 0.016,
\label{eq:SZlimit}
\end{equation}
since $y_{\text{true}}/\tau=k_BT_e/m_ec^2\approx 0.02$ for a
typical temperature of $10~$keV.
In a radially inhomogeneous universe smooth on sufficiently
large scales, the mean value of $v_r$ would
vary linearly with $z$ at small $z$.
Hence we should divide the above limit on $v_r/c$ by the typical
cluster redshift $z\approx 0.2$ to obtain a limit on inhomogeneity
$\approx 8\%$.
Since the temperature decrement is independent of distance,
multifrequency measurements at higher $z$ could---and probably
soon will---improve this limit substantially.

Peculiar velocities and inhomogeneities can also be constrained
by using
atomic and molecular excitation as a thermometer for the
CMB \cite{BW:68}.
One measures optical absorption from an excited level lying
$\sim k_B T_{\text{CMB}}$ above the ground state.
Clearly, it is important
to use systems in which collisional excitation is small or
negligible,
so that the observed excitation represents the angle-averaged
radiation temperature seen by the atomic
or molecular system, $\bar T_{\text{CMB},z}$.
Any discrepancy between this temperature and the redshifted
temperature of the local CMB can be explained by
a radial peculiar velocity:
\begin{equation}
\label{eq:discrep}
\frac{v_r}{c} \approx
\frac{(1+z)T_{\text{CMB},0}}{\bar T_{\text{CMB},z}} -1
\end{equation}
If $v_r$ is large or if it has a trend with $z$,
one has evidence of radial inhomogeneity.

Fine-structure lines of neutral carbon have been measured
in absorption (against a background quasar) at $z=1.776$,
with the result that $\bar T_{\text{CMB},z}= 7.4\pm0.8~$K, as
compared to $(1+z)T_{\text{CMB},0}= 7.58~$K \cite{S:94}.
Since the agreement is well within the errors, we can use the
errors to set a limit $\sim 0.8/7.58 = 11\%$.

Whether we consider scattering or absorption, the tests of this
section are sensitive primarily to dipole\footnote{In fact, by
analyzing the two photon polarizations separately in the cluster
test, one could measure both the dipole and one linear
combination of quadrupole moments.} anisotropies in the CMB
as it might be seen by distant observers.
For a cluster or absorption system at a given redshift $z$, the
measured temperature difference is linearly proportional to the
strength of the dipole seen at $z$.
The test described in the next section is sensitive to distant
anisotropies of all multipoles, but at second rather than first
order.

\section{Spectral distortions by diffuse scattering}
\label{sec:scatt}

Sufficiently strong radial
inhomogeneity at $z\sim 1000$ would produce a noticeable
spectral distortion because of the finite thickness of the
recombination surface \cite{Pe:93}.
This would measure inhomogeneities on a comoving scale only
$\sim 10^{-2}$ of the present horizon.

A spectral distortion sensitive to much larger scales could arise
from scattering by plasma associated with Ly$\alpha$ clouds and
a possible intercloud medium.
Absorption lines in quasar spectra reveal the presence
of diffuse, probably intergalactic, clouds at $z\sim 2-4$
containing small amounts of atomic hydrogen.
Physical considerations indicate that the hydrogen must be
predominantly ionized, and it is plausible that the ionized
intergalactic medium contains a significant fraction
$f_{\text{IGM}}\sim 1$ of the hydrogen indicated by cosmic
nucleosynthesis arguments
\cite{Pe:93,PR:93}, namely \cite{WSSOK}
\begin{equation}
\label{eq:nH}
\bar n_H\approx 0.76\bar n_B\approx
(1.1\pm 0.2)\times 10^{-7}\text{cm}^{-3}.
\end{equation}
As discussed in Sec.\ (\ref{sec:discuss}), the luminous parts
of galaxies account for only a
fraction of the $\bar n_H$ cited above, so that most of
the baryons in the universe must be sequestered in some form
other than visible stars.
Although an ionized IGM is not the only possible hiding place
for these baryons, it is a plausible one because hot gas
accounts for most the baryonic mass in Xray-emitting clusters of
galaxies \cite{WNEF,clusters}, and because searches sensitive
to local neutral hydrogen have found it in amounts much smaller
than Eq.\ (\ref{eq:nH}) \cite{WSGG}.

If the ionized IGM has persisted from $z_{\rm ion}=4$
to the present, then its total optical depth is
$\tau_{\text{IGM}}\approx 10^{-2}f_{\text{IGM}}$ if the present
age of the universe is $13$Gyr.
(With this choice of $z_{\rm ion}$, $\tau_{\text{IGM}}$ is
almost independent of $\Omega$ for $\Lambda=0$.)
On this assumption, about $1\%$ of the CMB photons that we
observe in any direction have been scattered at least once.
To the extent that the electrons are cold ($k_BT_e\ll m_ec^2$),
these scatterings have negligible effect on the CMB spectrum
in a homogeneous universe, but they will produce a slightly
nonthermal spectrum in a radially inhomogenous universe.

To make this discussion more quantitative, we consider scattering
in a spherically-symmetric but radially inhomogenous
matter-dominated universe \cite{Tolman}.
Since we cannot consider all possibile forms of inhomogeneity,
we shall adopt the following
simplifying assumptions and hope that our results are
representative of more difficult cases:

\begin{description}
\item[(S.1)] The model departs only slightly from a homogeneous
matter-dominated Einstein-de Sitter universe,
so that we may use first-order perturbation theory.
\item[(S.2)] The inhomogeneities are growing adiabatic
perturbations.
\item[(S.3)] Recombination occurs instantaneously at a fixed value
of the local temperature $T(r,t)=T_{\mbox{rec}}$.
\item[(S.4)] After recombination, the universe is sufficiently
optically thin that CMB photons scatter at most once before
reaching us, and most photons do not scatter at all.
\item[(S.5)] The photon energy is not changed in the local
matter rest frame by the scattering process.
\end{description}

Assumptions (S.1) and (S.2) are arbitrary and could be relaxed, but
they permit an easy Sachs-Wolfe treatment.
Concerning (S.3), the actual width of the recombination
epoch has been calculated to be $\Delta t/t\approx 0.2$
\cite{Pe:93}.
but to treat this epoch properly would require a dynamical analysis
of the interaction between matter and inhomogeneous radiation.
Assumption (S.4) is probably justified.
The possibility that the universe was  reionizeed early enough to
produce a substantial optical depth has been much discussed,
but this appears unlikely because  CMB anisotropies are now seen
on degree scales \cite{WSS:94}.
As we shall see, the spectral distortions produced by
inhomogeneity would be indistinguishable from those caused by a
hot intergalactic medium in a homogeneous universe.
The two effects are additive and cannot be made to cancel,
so assumption (S.5) is conservative.

We look out towards the recombination epoch along past-directed
null geodesics.
According to assumption (S.4), most of the CMB photons we receive
never scattered on their way to us, so these photons sample the
recombination epoch on a sphere of comoving radius $r_{\text{d}}$.
(``d'' for ``direct''.)
Because of isotropy, all photons reaching us from this sphere have
been drawn from a Planck distribution with a common temperature
and have suffered the same redshift, so their spectrum is
completely thermal.

A minority of photons have scattered once.
Consider all such photons that have scattered towards us through
angle $\psi$ at some epoch $t_{\rm s}$, where
$t_{\rm rec}<t_{\rm s}<t_{\rm now}$.
Traced backwards from $t_{\rm s}$, the paths of these photons
intercept the recombination era on a common sphere of radius
$r_{\text{i}} (t_{\text{s}},\psi) < r_{\text{d}}$
(``i'' for ``indirect'').
Because of assumption (S.2), photons from the spheres
$r_{\text{d}}$ and $r_{\text{i}}$ were drawn from the same Planck
distribution.
In a homogeneous universe, photons from both spheres would suffer
the same redshift, $(1+z_{\rm rec})$, and because of this and
assumption (S.5), the CMB  spectrum would be unaffected by
scattering.
In an inhomogeneous universe, however, there is a first-order
perturbation in these redshifts:
\begin{eqnarray}
\label{eq:dzeqn}
&\zeta &\equiv \delta\ln(1+z) =
\case{1}{3}\left\{ \phi(0) -\phi[r(\eta_{\rm rec})]\right\} \\
&+&\frac{1}{3}\left\{(1-\cos\psi)\eta_s
\frac{d\phi}{dr}[r(\eta_{\rm s})]
+\eta_{\rm rec}\left(\frac{dr}{d\eta}
\frac{d\phi}{dr}\right)_{r(\eta_{\rm rec})}\right\} \nonumber
\end{eqnarray}
Here $r=\sqrt{x^2+y^2+z^2}$ is the comoving radius in the
unperturbed Einstein-de Sitter metric,
\begin{equation}
\label{eq:metric}
ds^2 = dt^2 - a^2(t)(dx^2+dy^2+dz^2),
\end{equation}
$\eta$ is the arc parameter,
\begin{equation}
\label{eq:etadef}
\eta(t)\equiv \int\limits_0^t\frac{dt'}{a(t')},
\end{equation}
$\phi(r)$ is the potential fluctuation associated with the
perturbation in the mass density,
\begin{equation}
\label{eq:Poisson}
\phi(\vec r)= - a^2(t)
G\int\frac{d^3\vec r^{\,\prime}}{|\vec r-\vec r^{\,\prime}|}
\delta\rho(\vec r^{\,\prime},t),
\end{equation}
and the path of the photon is
(for $\eta_{\text{rec}}\le\eta_{\text{s}}\le\eta_0$)
\begin{equation}
\label{eq:rofeta}
r(\eta) = \cases{
[(\eta_0-\eta_s)^2 +(\eta_s-\eta)^2 & \cr
{}~+ 2(\eta_0-\eta_s)(\eta_s-\eta)\cos\psi]^{1/2} &
$\eta\le\eta_s$,\cr
\eta_0-\eta &  $\eta\ge\eta_s$, \cr}
\end{equation}
where $\eta_0\equiv\eta(t_0)$ is the present epoch,
$\eta_s\equiv\eta(t_s)$ is the epoch of scattering,
and $\psi$ is the scattering angle.
The potential $\phi(r)$ defined by (\ref{eq:Poisson}) is
time independent, since $\delta\rho\propto a^{-2}$ for the growing
mode [$\delta\rho/\rho\propto a\propto t^{2/3}$].
That the redshift perturbation can be written in terms of such
a potential is the fundamental result of Sachs and Wolfe \cite{SW}.

The first two terms contributing to the redshift perturbation
(\ref{eq:dzeqn})
reflect the difference in potential between the origin and
destination of the photon.
The remaining terms are Doppler shifts: they  arise from
the peculiar velocities of the matter with respect to the
unperturbed background Einstein-de Sitter cosmology.
These velocities
are produced by the peculiar gravitational acceleration $-d\phi/dr$
acting over arc-parameter ``time'' $\eta$.
In our case, the peculiar velocities are radial, so the Doppler
shift is proportional to $dr/d\eta$, which is the cosine of
the angle between the photon momentum and the radial direction.
The Doppler shift at the scattering epoch depends upon the change
in that cosine, whence the $(1-\cos\psi)$ factor.
Since $\eta_{\text{rec}}/\eta_0=(1+z_{\text{rec}})^{-1/2}\approx
10^{-1.5}\ll 1$, we simplify Eqs.\ (\ref{eq:dzeqn}) and
(\ref{eq:rofeta}) by replacing $\eta_{rec}$ with $0$.

The CMB spectrum seen by an observer at $r=0$ is a weighted sum of
redshifted Planck functions:
\begin{equation}
\label{eq:convol}
I_\nu(r=0) = \int B_\nu(e^{-\zeta}T_0)dP(\zeta)
\end{equation}
where $B_\nu(T_0)$ is a Planck function at the present-day CMB
temperature, $T_0$ $\equiv T_{\text{rec}}/(1+z_{\text{rec},0})$,
and $P(\zeta)$ is the probability distribution for $\zeta$.
The probability density $dP/d\zeta$ consists of two parts: a
delta function of area $e^{-\tau_{\text{IGM}}}$ at the value of $\zeta$
for unscattered photons [computed from (\ref{eq:dzeqn}) by setting
$\psi=0$]; and a continuous part of total area
$1-e^{-\tau_{\text{IGM}}}\approx\tau_{\text{IGM}}\ll 1$
representing the once-scattered photons.

It is clear that if $\zeta$ were the same along all paths, the
spectrum $I_\nu(0)$ would remain thermal but would have temperature
$e^{-\zeta}T_0$ instead of $T_0$.
Thus nonthermal distortions depend upon the differences in $\zeta$
among scattered paths.
This can be demonstrated formally by expanding $B_\nu(T)$ in
equation (\ref{eq:convol}) as a Taylor series in $\ln T$ about the
point $\ln T=\ln T_0 -\langle\zeta\rangle$, with the result
\begin{eqnarray}
\label{eq:series}
I_\nu(0)&=& B_\nu(T_0 e^{-\langle\zeta\rangle})\nonumber \\
&+& \left[\sum\limits_{n=2}^\infty
\langle(\zeta-\langle\zeta\rangle)^n\rangle
\left(T\frac{\partial}{\partial T}\right)^n B_\nu(T)
\right]_{T_0 e^{-\langle\zeta\rangle}}
\end{eqnarray}
The moments of $\zeta$ are
\begin{equation}
\label{eq:moments}
\langle\zeta^n\rangle\equiv\int\zeta^n dP(\zeta).
\end{equation}
If the width of the redshift distribution $P(\zeta)$ is small, as
it is when $|\phi(r)|, ~|r\phi'(r)|\ll c^2$, we do not need to go
beyond the second-derivative term in the expansion
(\ref{eq:series}).
So we have only to compute the variance in the log redshift.
The unscattered paths contribute to $\langle\zeta\rangle$ but not
to the variance $\langle(\zeta-\langle\zeta\rangle)^2\rangle$.
So we may compute modified moments, marked by a prime, by averaging
over the scattered paths only:
\begin{equation}
\label{eq:smoments}
\langle\zeta^m\rangle' = \int\limits_{\eta_{\text{rec}}}^{\eta_0} d\eta_s
\frac{d\tau}{d\eta_s} \int\limits_{-1}^{+1} d\cos\psi
\,\case{3}{8}(1+\cos^2\psi)\zeta^m(\eta_s,\psi),
\end{equation}
where $\zeta(\eta_s,\psi)$ is the function (\ref{eq:dzeqn}).

The differential optical depth is
\begin{eqnarray}
\label{eq:opticaldepth}
\frac{d\tau}{d\eta_s}=&\sigma_T& n_e(t)c\frac{dt}{d\eta_s}=
3\sigma_T n_e(t_0)\frac{ct_0}{\eta_0} \nonumber \\
&\times&\cases{(\eta_s/\eta_0)^{-4} &if $\eta_s\ge
\eta_{\text{ion}}$, \cr
0 & if $\eta_{\text{rec}}<\eta_s <\eta_{\text{ion}}$.\cr}
\end{eqnarray}
Following the discussion above, we have taken the comoving density
of electrons to be constant from the present back to a redshift
factor $1+z_{\text{ion}}= (\eta_{\text{ion}}/\eta_0)^{-2}$.
The total optical depth is (assuming full ionization of $^4$He)
\begin{eqnarray}
\label{eq:totaltau}
\tau_{\text{IGM}}&=& \sigma_T n_e(t_0)c t_0[(1+z_{\text{ion}})^{3/2}-1]
\nonumber \\
&\approx& 1.0\times 10^{-3}f_{\text{IGM}}[(1+z_{\text{ion}})^{3/2}-1].
\end{eqnarray}

To compute the variance of $\zeta$ and hence the
distortion (\ref{eq:series}), we must assume a functional form
for the perturbed potential $\phi(r)$.
Rather arbitrarily, we choose
\begin{equation}
\label{eq:phiform}
\phi(r)= \phi_0\cos(\omega r),
\end{equation}
in which $\phi_0$ is a normalization and $2\pi/\omega$ is an adjustable
comoving radial wavelength.
A general spherically-symmetric linear perturbation could be written
as a superposition of Fourier components of this form.
To fix the meaning of $\omega$, $a(t)$ will be scaled so
that $\eta_0=1$ [$a(t_0)= 3 t_0$].
Therefore $r=1$ is the present horizon, and $a(t_0)=3t_0$.
After substitution of equations (\ref{eq:dzeqn}) and (\ref{eq:phiform})
into the formula for the moments (\ref{eq:smoments}), the integrations
over $\psi$ and $\eta_s$ can be expressed in closed but lengthy form,
which we omit.

The variance $\langle\zeta^2\rangle-\langle\zeta\rangle^2\equiv
\langle\Delta\zeta^2\rangle$ is clearly proportional to
$\phi_0^2$, but it vanishes for certain functional forms of $\phi(r)$.
In the limit $\omega\to0$, $\phi(r)$ becomes constant and
has no affect on the observed spectrum [cf. equation (\ref{eq:dzeqn})].
Even a quadratic potential $\phi(r)=\phi_0\cdot[1-(\omega r)^2/2]$
would spoil neither the thermality of $I_\nu(0)$ nor the homogeneity
of the geometry: such a perturbation corresponds
to a slightly non-flat Friedmann model.
It follows that for $\omega\ll 1$, $\langle\zeta^2\rangle\propto
\omega^8\phi_0^2$.

One coordinate-independent way to characterize
the sensitivity of the spectrum (\ref{eq:series})
to the degree of radial inhomogeneity is to
compare the spectral distortion at $r=0$ with
the {\it anisotropy} seen by a typical {\it non-central} observer.
In the limit $\tau_{\text{IGM}}\to0$,  an observer at $r_{\rm obs}>0$ and
$\eta=\eta_0$ sees
the CMB as thermal in every direction, but the temperature
varies with angle according to
\begin{equation}
\label{eq:angdT}
\frac{\delta T}{T}(\theta) = \frac{1}{3}\left\{
\phi[r_{\text{rec}}(\theta)]-\phi(r_{\rm obs}) +
\eta_0\cos\theta\frac{d\phi}{dr}(r_{\rm obs})\right\}
\end{equation}
apart from a constant.
Here $\theta$ is measured with respect to the radial direction, and
\begin{equation}
\label{eq:rrec}
r_{\text{rec}}(\theta)= [r_{\rm obs}^2 +\eta_0^2 +2\eta_0 r_{\rm obs}
\cos\theta]^{1/2}
\end{equation}
is the locus of the non-central observer's horizon.
We calculate the variance
\begin{eqnarray}
\label{eq:anvar}
\left(\frac{\Delta T}{T}\right)^2_{r_{\text{obs}}} &\equiv &
\int\limits_{-1}^{+1}d\cos\theta
\left[\frac{\delta T(\theta)}{T}\right]^2 \nonumber \\
&-&\left[\int\limits_{-1}^{+1}d\cos\theta
{}~\frac{\delta T(\theta)}{T}\right]^2.
\end{eqnarray}
When $\omega\gg 1$, $(\Delta T/T)^2$ can
oscillate rapidly with $r_{\rm obs}$,
so we compute a smoothly-tapered radial average:
\begin{equation}
\label{eq:smooth}
\left\langle\left(\frac{\Delta T}{T}\right)^2\right\rangle\equiv
\int\limits_0^1 dr_{\text{obs}}\left(\frac{\Delta T}{T}
\right)^2_{r_{\text{obs}}}
\case{1}{2}\sin^2(\pi r_{\rm obs}).
\end{equation}

Finally, we define the ``normalized'' spectral distortion $\hat y$
by
\begin{equation}
\label{eq:ynorm}
\tau_{\text{IGM}}\hat y \equiv \frac{\langle\Delta\zeta^2\rangle}{
\langle(\Delta T/T)^2\rangle}
\end{equation}
Since both the numerator (the spectral distortion at the center
of the universe) and the denominator
(the typical angular temperature variance off center) are
proportional to $\phi_0^2$, the quantity
$\hat y$ is independent of the amplitude of the potential
fluctuations in the linear regime.
We have also scaled the total optical depth (\ref{eq:totaltau})
out of $\hat y$.
However, $\hat y$ does depend somewhat on $\omega$ and
$z_{\text{ion}}$.

Table \ref{table1} shows some representative values of $\hat y$.
The notation ``$0^+$'' under the heading for $\omega/2\pi$ denotes
the limit as $\omega\to 0$.
One sees from the Table that $\hat y$ is essentially independent of
$\omega$ if there are two or more cycles within the horizon
($\omega/2\pi\gtrsim 2$), but $\hat y$ decreases sharply at smaller
$\omega$.
Thus the spectral distortion is relatively insensitive to density
perturbations  that are well fit by a quadratic function of radius
(corresponding to a quartic potential, since
 $\delta\rho/\rho\propto\eta^2\nabla^2\phi$).

The value of $\hat y$ decreases noticeably if $z_{text{ion}}\gg 1$.
In that case,
since the differential optical depth $d\tau\propto \sqrt{1+z}$,
most of the scatterings occur very early, at
$z\sim z_{\text{ion}}$,
when the photons have not moved far from their positions at
recombination.
If $\omega$ is moderate,
the potential $\phi$ is nearly constant over this small range
in $r_i$.
On the other hand,
if $\omega\gg 1$, the redshift perturbation (\ref{eq:dzeqn})
is dominated by the Doppler term $\eta_s d\phi/d\eta_s$, which
decreases with scattering epoch $\eta_s\propto(1+z_s)^{-1/2}$.
[For $\omega\gg 1$, the denominator of (\ref{eq:ynorm}) is also
dominated
by Doppler shifts---those of the noncentral observers
themselves---but these shifts are evaluated at the present epoch
and are not suppressed by the $(1+z_{s})^{-1/2}$ factor.]
So $\hat y$ is small for large $z_{\text{ion}}$,
regardless of $\omega$.

We have scaled the total optical depth
$\tau_{\text{IGM}}$ out of $\hat y$, however, so that
the observed distortion is proportional to $\tau_{\text{IGM}}\hat y$.
For $z_{\text{ion}}=2,~4$, and $10$, equation (\ref{eq:totaltau})
predicts $\tau_{\text{IGM}}\approx 0.004 f$,
$0.01 f$, and $0.04 f$, respectively; at larger $z_{\text{ion}}$,
the approximation of single scattering begins to break down.
At any rate, for a fixed nonzero amplitude $\phi_0$ and for
$\omega>0$, the observed spectral distortion tends to increase
with $z_{\text{ion}}$ despite the decrease in $\hat y$.

We have adopted the notation ``$\hat y$'' because the spectral
distortion produced by radial inhomogeneity has the same form as
the distortion arising from inverse compton scattering provided
that $T_\gamma\ll T_e\ll m_e c^2/k_B$, where $T_\gamma$ and $T_e$
are the photon and electron temperatures.
Quite generally, a mixture of Planck functions at slightly
different temperatures is indistinguishable from a slightly
comptonized spectrum (\cite{D:91}).
The correspondance in Eq.\ (\ref{eq:series}) is
$\langle\zeta\rangle\to -3y$ and
$\langle\Delta\zeta^2\rangle\to y$; higher-order moments of
$\zeta$ are negligible if $y\ll 1$.

This means that we can translate published upper limits on
comptonization of the CMB into limits on radial inhomogeneity.
The limit reported by the COBE collaboration is
$y < 2.5\times 10^{-5}$ (\cite{Ma:94}).
Hence the limit on inhomogeneity as defined by (\ref{eq:smooth}) is
\begin{eqnarray}
\label{eq:limit}
\overline{\left(\frac{\Delta T}{T}\right)} &\le & \left(
\frac{y}{\tau_{\text{IGM}}
\hat y(\omega,z_{\text{ion}})}\right)^{1/2}\nonumber \\
&\lesssim & 0.05 f^{-1/2}.
\end{eqnarray}
In the final numerical form, we have taken $\hat y\approx 1$ as a
typical value from Table 1, and we have assumed $z_{\text{ion}}=4$.
As Table \ref{table1} shows, $\hat y$
is much smaller and our limit correspondingly weaker if
$\omega<\pi$.

\section{Discussion}
\label{sec:discuss}

We have discussed three tests of the large-scale radial
homogeneity of the universe.
All three involve possible distortions or variations in the CMB
spectrum. There is, however, a fourth test that is more widely
recognized than any of these: big-bang nucleosynthesis of the
light elements (henceforth BBN).

The standard theory of BBN predicts the
primordial abundances of the light elements $^1$H, $^2$H, $^3$He,
$^4$He, and $^7$Li in terms of a single parameter,
$\eta$, the number of baryons per CMB photon.
(We ignore complications such as a nonstandard number of neutrino
flavors.)
{}From the observed {\it relative} abundances of these elements in
the local universe, it appears that \cite{WSSOK}
\begin{equation}
\label{eq:etarange}
2.8\le\eta_{10}\le 4.0,~~~\eta_{10}\equiv 10^{10}\eta.
\end{equation}

Measuring abundance ratios such as $N(^2\text{H}/N(^1\text{H})$
in cosmologically distant systems tests radial
homogeneity without reference to the CMB or to BBN,
since it is sensitive to inhomogeneities in whatever processes create
and destroy these elements.
Since such measurements have only just begun to come in during the
past year \cite{SCHR},
and since a consensus has not yet been reached,
this is not the fourth test we referred to above.
Nevertheless it may give very interesting results in the near
future.

The value of $\eta$ constrained by (\ref{eq:etarange}) pertains to
photons that resided here at the time of nucleosynthesis.
Those photons are no longer with us: they have been streaming away
since recombination and are now almost at the horizon.
On the hypothesis of homogeneity, however, the CMB photons
seen today stand as proxies for those long-gone photons.
In particular, we may evaluate $\eta$ using the present-day density
of CMB photons in our vicinity, $n_\gamma\propto T_{\text{CMB}}^3$.
Hence the local mean density of baryons should be
\cite{WSSOK}
\begin{equation}
\bar n_B= 20.3 T_{\text{CMB}}^3\eta~\text{cm}^{-3}\approx
(1.4\pm 0.3)\times 10^{-7}~\text{cm}^{-3},
\label{eq:nB}
\end{equation}
where we have taken $T=2.726~K$ (\cite{Ma:94})
and adopted the range (\ref{eq:etarange}) for $\eta$.
To the extent that $\bar n_B$ can be measured and compared with
this prediction, one tests the {\it spatial} constancy of $\eta$.
Under the assumptions {\bf (S.1)-(S.5)} of Sec.\ \ref{sec:scatt},
we have
\begin{equation}
(n_B^{\rm meas.}/n_B^{\rm pred.})~-1 = \phi(0)
-\phi(r=\eta_{\text{rec}}),
\label{eq:nBlimit}
\end{equation}
provided both sides of this equation are small compared to unity.
To be competitive with the limits presented in
Secs.\ \ref{sec:therm}-\ref{sec:scatt},
one should measure $\bar n_B$ to $\sim 30\%$ or better.

It seems that $\bar n_B$ has not yet been measured to the
required accuracy.
It is estimated that the luminous parts of galaxies
account for a mean mass density
$\approx 5\times 10^{-8}e^{\pm 0.3} h^2 m_H~\text{cm}^{-3}$,
where $h\equiv H_0/(100~{\rm km~s^{-1}~Mpc^{-1}})$,
based on the observed luminosity density and an assumed
mass-to-light ratio $M/L_B = 12 M_{sun}/L_{sun}$ \cite{Pe:93}.
Since this ratio counts all of the mass within the Holmberg
radius, it may include some nonbaryonic dark matter;
and probably $h<1$.
Thus the visible parts of galaxies fall short of the
nucleosynthetic prediction (\ref{eq:nBlimit}) by a factor of
at least $3$.
In some clusters of galaxies, hot Xray-emitting gas
increases the total baryonic $M/L_B$ by a factor
$5.6$ to $16$, depending on $h$ \cite{WNEF}.
In summary, while the observed baryon density is
perhaps consistent with Eq.\ (\ref{eq:nB}),
the bookkeeping is not yet
accurate enough to yield a $10\%$ limit on the
radial homogeneity of the CMB temperature.

\acknowledgements

The author thanks Bohdan Paczy\'nski for posing the question that
this letter attempts to answer, and Ruth Daly,
Puragra Guhathakurta, Tod Lauer, J.P. Ostriker,
P.J.E. Peebles, David Spergel, and Ralph Wijers
for helpful discussions.
Special thanks are due the Packard Foundation for its generous
financial support during the germination of this work.

\begin{table}
\caption{Normalized spectral distortion $\hat y$.
\label{table1}}
\begin{tabular}{llll}
$\omega/2\pi$ & $z_{\text{ion}}=2$ & $z_{\text{ion}}=4$ &
$z_{\text{ion}}=100$ \\
\tableline
$0^+$ & $0.040$ & $0.029$ & $0.0027$ \\
$0.5$ & $0.16$  & $0.11$  & $0.0076$ \\
$1.0$ & $1.31$  & $0.98$  & $0.026$  \\
$2.0$ & $2.31$  & $1.23$  & $0.056$  \\
$10.$ & $2.18$  & $1.49$  & $0.10$   \\
$10^3$& $2.20$  & $1.53$  & $0.11$   \\
\end{tabular}
\end{table}

\end{document}